\documentclass[preprint]{aastex}
\usepackage{epsf}
\usepackage{natbib}
\usepackage{subfigure}
\usepackage{rotating}
\usepackage{epstopdf}

\newcommand{\feh}{\hbox{$ [{\rm Fe}/{\rm H}]$ }}
\newcommand{\fehc}{\hbox{$ [{\rm Fe}/{\rm H}]$}}
\newcommand{\dellpc}{\hbox{$ {\rm \Delta logP}$}} 
\newcommand{\dellp}{\hbox{$ {\rm \Delta logP}$ }}

\begin{document}

\title{An Oosterhoff Analysis of the Galactic Bulge Field RR Lyrae stars: Implications
On Their Absolute Magnitudes}

\author{Andrea Kunder and Brian Chaboyer} 
\affil{Dartmouth College, 6127 Wilder Lab, Hanover, NH 03755}
\affil{E-mail: Andrea.M.Kunder@Dartmouth.edu and Brian.Chaboyer@dartmouth.edu}

\begin{abstract}
We present an analysis of the period-$V$-amplitude plane for RR0 Lyrae 
stars (fundamental mode pulsators) with ``normal'' light curves in the 
bulge using the MACHO bulge fields.  Although bulge globular 
clusters have RR Lyraes that divide into two reasonable distinct groups
according to the average period of the RR0 Lyraes \citep{ooster39}, 
there is no evidence of a gap between Oosterhoff I and II stars in the 
bulge field star sample.  The majority of the 
bulge RR0 Lyrae field star population have a difference in period compared
to the Oosterhoff I cluster M3 (\dellpc) that is shifted
by about 0.02 days with regard to the Milky Way Oosterhoff I population,
and the sample includes stars with \dellp $>$ 0.06 days, a characteristic
hardly seen in Milky Way globular clusters.  The metal-rich RR0 Lyrae stars in
the Galactic bulge sample have \dellp values on the other side of the spectrum
as those in the metal-rich globular clusters NGC 6388 and NGC 6441.
We find that the $V$-amplitude for a given period is a 
function of \dellpc, and not of metal abundance, similar to the result
found by \citet{clement99} for RR Lyrae stars in Milky Way globular
clusters.  A comparative study of the bulge field
stars with similar metallicities but different Oosterhoff types is carried 
out.  Bulge field RR0 Lyrae variables with \dellp values similar to 
Oosterhoff II clusters 
are about 0.2 mag brighter than RR0 Lyrae variables with \dellp similar to 
Oosterhoff I clusters.  Reliance upon a single $\rm M^{\mathrm{RR}\,}_{V}$
-\feh relation
may be inappropriate when considering populations with different 
\dellpc.  
\end{abstract}

\section{Introduction}
RR Lyrae stars are the most popular distance indicators for old, low mass 
stars \citep[see, e.g.,][]{carretta00,walker00}.  The knowledge of their absolute 
magnitude and its possible dependence 
on metallicity allows the determination of basic astronomical parameters such 
as distance to the galactic center and Local Group galaxies, as well as
distances and ages of the Milky Way globular clusters.  These have a connection 
with the Hubble constant and with the conditions of the Milky Way's Galaxy formation.

It was shown by \citet{ooster39} that globular clusters (GCs) divide 
into two groups, based on the pulsational properties of their fundamental 
mode pulsating RR0 Lyrae stars.  Later it was found that these groups
could also be separated by metallicity.  The Oosterhoff I (OoI) 
type GCs have an average RR0 Lyrae pulsational period of 0.55 days
and metallicities that are more metal rich than $-$1.5 dex.  The Oosterhoff
II (OoII) type GCs have an average RR0 Lyrae pulsational period of 0.65 days
and are more metal poor than $-$1.5 dex.  There is a gap at the two period
groups separated by a mean period ratio of log 0.65/0.55 = 0.073.  This is now
termed the Oosterhoff dichotomy.  No physical reason 
for this dichotomy in periods is known to date, but a difference in mean 
luminosity between the two groups of about 0.2 mag could explain the 
observations \citep{sandage06}.  Temperature differences alone between the 
two Oosterhoff period groups are not sufficient to 
account for this luminosity difference \citep{sandage58}.

It has long been customary to express the luminosities of RR Lyrae variables by
\begin{equation}
\label{mvfe}
M_V = a + \feh b.
\end{equation}
However, there has been much dispute as to the values of $a$ and $b$.  
The question of the linearity of the equation is also now being investigated.  
The evidence for non-linearity is from (a) theoretical models of the
horizontal-branch (HB) (both at zero age and in an evolved state), (b) the 
pulsation equation using the observed input parameters of mass-\feh and 
log $\rm T_e$ (from colors) as functions of metallicity relations, (c) and 
from semiempirical use of observational data \citep{sandtamm06}.   For 
example, Bono, Caputo, \& Di Criscienzo (2007) use RR Lyrae in the
Milky Way show the nonlinearity 
(see their Fig. 16) over the entire range of [Fe/H] from 0 to $-$2.5 dex.

The derived zero points of the nonlinear 
$M^{\mathrm{RR}\,}_{V} -\feh$ relation depends on the
investigations and methods used.  Some lead to the so-called long 
RR Lyrae scale that gives $\langle M_{V}\rangle$ $\sim$ +0.52 
at [$\mathrm{Fe}\,/ \mathrm{H}\,$] = $-$1.5 dex and some lead to
the short scale that gives $\langle M_{V}\rangle$ $\sim$ 
+0.72 at [$\mathrm{Fe}\,/ \mathrm{H}\,$] = $-$1.5 dex.
Gaining an understanding of the Oosterhoff dichotomy will likely aid in 
understanding the absolute magnitude calibration of RR Lyrae stars.
In this paper we look at the Oosterhoff properties of bulge field stars and present 
evidence that even nonlinear \feh relations
cannot fully account for finding their absolute magnitudes.

Adding to the Oosterhoff puzzle, Local Group Dwarf spheroidal galaxies 
(dSph) and LMC globular clusters have RR0 Lyrae variables with properties 
that fill the gap in a plot of \feh versus fundamental mode period 
as illustrated by \citet{siegel00},  \citet{pritzl04} and \citet{catelan05}. 
Studying the Oosterhoff properties of GCs in both our Galaxy
and those belonging to the dwarf companions of the Milky
Way aid in identifying the building blocks of the Galactic halo.
Identifying the origin of the Oosterhoff dichotomy is hence 
helpful in exploring the formation of the Milky Way galaxy.  For example, if
the \citet{searle78} Galaxy formation scenario (hierarchical merging and 
continual accretion of lower mass protogalactic fragment formation scenario) 
is dominant, then one would imagine that the RR Lyrae variables in the Local 
Group dSph's and the RR Lyrae variables in the LMC globular
clusters would have similar properties to that of the Milky Way RR Lyrae stars 
(see, e.g., Catelan 2007).

One explanation for the Oosterhoff dichotomy was put forth by \citet{vanalbada73},
They suggested that the evolutionary scenario of RR Lyrae in OoI and OoII clusters
differs.  The horizontal-branch (HB) stars in the intermediate metallicity OoI
clusters begin their lives near the instability strip.  As they evolve they
increase their temperature, and hence fundamental pulsating RR0
Lyrae would develop into first-overtone RR1 variables. 
In contrast, the horizontal-branch stars in the more metal-poor OoII clusters
approach the instability strip from a different direction that those in
the intermediate metallicity clusters.  They begin their lives at higher temperatures
and evolve to lower temperatures.  The existence of an hysteresis zone
is suggested by \citet{vanalbada73}, so that depending on the direction of 
evolution, mode switching is delayed and occurs at different temperatures.  
Using convective pulsating models as well as observations
of horizontal-branch stars in two different Oosterhoff GCs, 
Bono, Caputo \& Marconi (1995) were able to show that the transition 
between RR0 and RR1 RR Lyrae supports the hysteresis mechanism.

The \citet{vanalbada73} explanation thus suggests (1) OoI clusters are evolving 
red to blue and OoII clusters the other way, (2) that OoI clusters have
a smaller fraction of RR1 variables than OoII clusters, and (3) that
metallicity is the dominant characteristic that determines where on
the horizontal-branch core helium burning begins.  However, this explanation 
does not explain the absence of a 
dichotomy in other galaxies unless the range in metallicity in each 
galaxy is very small \citep{lee99}. 

Ultimately, the issue of the second parameter problem in the
horizontal-branch morphology could potentially account for the Oosterhoff
dichotomy.  Bluer and more extended HBs exist in the more metal-poor 
clusters while metal-rich ones have 
more of a red clump (Lee, Demarque, \& Zinn 1990).  
Yet as this is not always the case, there must be another factor, 
the second parameter, which affects HB morphology.  Two possibilities of this
second parameter could be helium enrichment through a variety of mechanisms 
\citep{sweigart98} and age.  Now with the discovery of metal-rich bulge
GCs that do not follow the period-metallicity 
trends of the other two groups \citep{pritzl01}, the second parameter
issue is of considerable importance to understand these ``Oosterhoff III"
GCs.

The Oosterhoff dichotomy was first seen in GCs, but there has 
been quite a bit of interest to see how field RR Lyrae stars pertain to the
Oosterhoff dichotomy.  Recently using RR0 Lyrae stars in the LONEOS-I,
SDSS-II Supernova Survey and All Sky Automated Survey (ASAS) two 
distinct OoI and OoII components were found by examining the RR Lyrae
period shift \citep{miceli08, delee08, szczygiel09}.  
The LONEOS-1 survey covers 1430 $\rm deg^2$ to a distance of 30 kpc 
and traces the ecliptic.  The SDSS-II Supernova Survey covers a 2.5$^\circ$ 
equatorial stripe ranging from $-$60 to +60 degrees in RA, and covers a 
distance of 3 to 95 kpc from the sun.  ASAS surveys the southern sky up 
to $\delta$ = +28 and reaches to distance of $\sim$ 5 kpc.

In contrast, the QUEST-I RR Lyrae
\citep{vivas04} and the NSVS RR Lyae \citep{kinemuchi06} did not show a clear 
OoII component.  The NSVS survey was recently shown to have incorrect 
$V$-amplitudes, which concealed the presence of two Oosterhoff components 
in their sample \citep{szczygiel09}.  The NSVS
survey spans a region of 5 kpc from the Sun and primarily 
covers the entire northern sky.  
The QUEST RR Lyrae survey covers 380 $\rm deg^2$ 
to a distance of 55 kpc, and covers equatorial regions of the sky.  We investigate here if the Oosterhoff 
dichotomy is present in the field Galactic bulge RR0 Lyrae stars.  This 
region is distinct from the LONEOS-I, QUEST, ASAS, SDSS-II and NSVS surveys.
Further, this region has metal-rich RR Lyrae stars, which, except for the 
recently discovered ``OoIII'' clusters mentioned above, are hardly seen
in the Milky Way Globular Clusters.

\section{Data}
We consider the RR0 Lyrae stars from the 
MACHO bulge fields with ``normal'' light curves \citep{kunder08}.
A ``normal'' light curve was first described by \citet{jk96} who 
used interrelations among the lower order Fourier parameters
to derive a set of equations for calculating Fourier amplitudes and
phase differences.  A star's light curve is considered ``normal'' if
the calculated values for all of these parameters are in good agreement with
the observed values and hence has a low deviation parameter.  
Since then, \citet{kovkan98} have updated the deviation parameter analysis,
and their values for deviation parameters are used in this analysis.  
The stars from  \citet{kundchab08} have published photometric 
$\rm [Fe/H]$ metallicities determined from their Fourier 
decomposition parameters as well as reddening estimates as 
determined from their $\rm (V-R)$ color at 
minimum $V$-band light.

Only ``normal'' RR Lyrae variables are
used because the Period-Amplitude ($PA_V$) plane, which is often used 
to distinguish between Oosterhoff type I and Oosterhoff
type II globular clusters, becomes contaminated by Blazhko variables.
Blazhko variables are RR Lyrae stars with non-repeating light curves, and hence
their amplitudes vary over timescales longer than the basic pulsation period.
If not identified, they introduce scatter into the $\rm PA_V$ diagram.  
Using those stars with ``normal" light curves significantly reduces a major contributor of 
contamination in the $\rm PA_V$ diagram.  

\begin{figure}[htb]
\includegraphics[width=16cm]{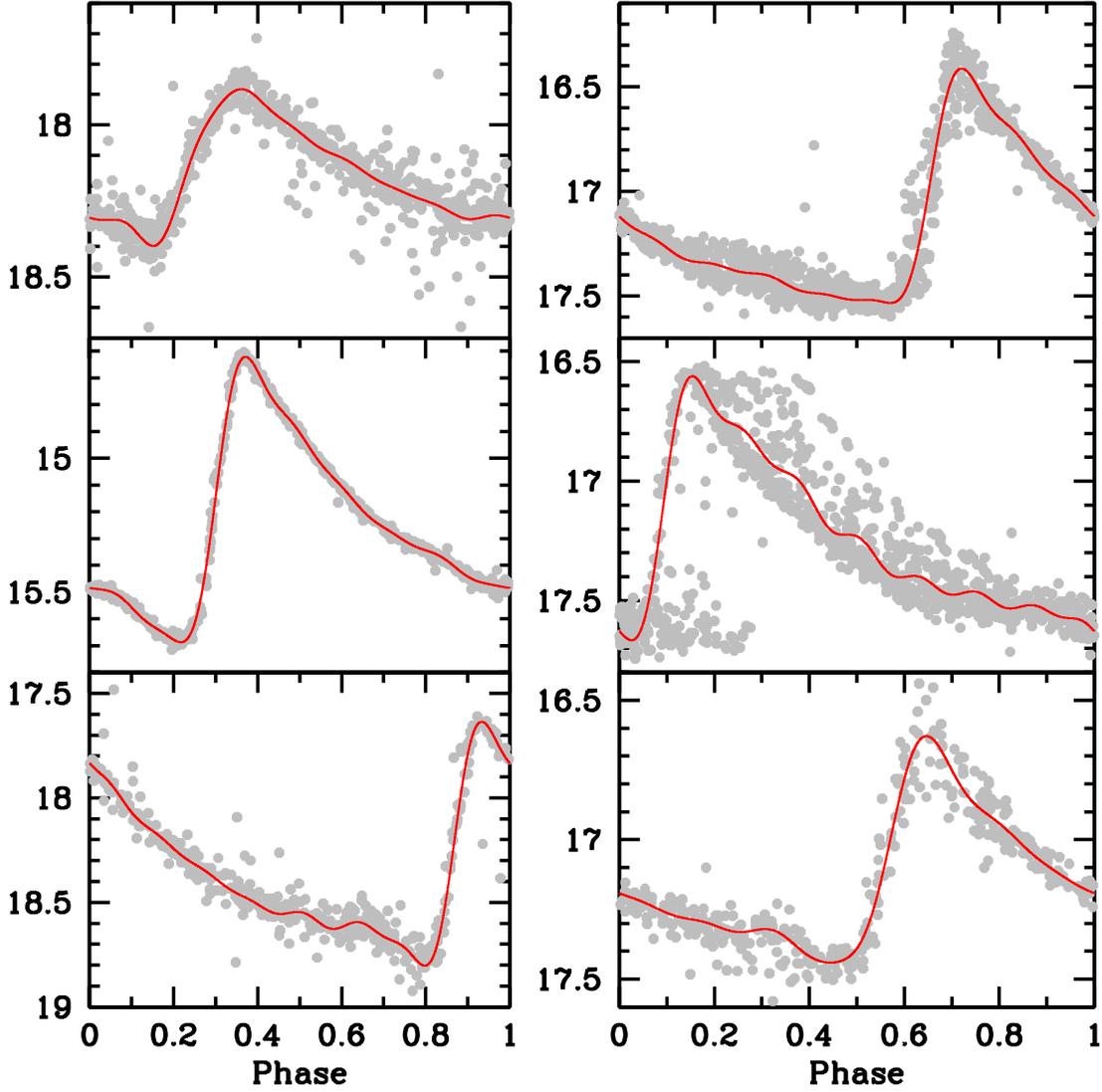}
\caption{\textit{Left:} Three of the 1323 high quality MACHO bulge light curves.  From top to 
bottom these stars are MACHO 101.20909.1342, 110.22316.115 and 403.47909.1408. \textit{Right:}
Three of the 1028 standard MACHO bulge light curves.  From top to bottom
these stars are MACHO 104.20388.631, 104.21164.904 and 109.20114.714.
The $V$-band magnitudes are plotted as function of phase using the periods from
the MACHO database.  The white lines represent the 8th order Fourier fit to the data.
The error bars are omitted for clarity.
\label{f1}}
\end{figure}

To further asses the quality of the light curves, all the ``normal" light curves 
were visually inspected and the stars with the most high quality light curves were 
selected.  
Stars that exhibit the Blazhko effect have especially notable scatter at the 
height of maximum of their light curves, and hence, we particularly checked
that the scatter around maximum light was minimal, or at least consistent with the
scatter of the light curve.  This resulted in a sample of 1323 high quality light curves.
Examples of three typical high quality and three typical standard light curves
are shown in the left and right panels of Figure~\ref{f1}.

We now discuss the reliability of a 
MACHO RR Lyrae stars' period and $V$-amplitude measurement; hence the 
reliability of a MACHO RR Lyrae's location on a $\rm PA_V$ diagram can be 
assessed.  An RR0 Lyrae stars' period is perhaps the most well determined parameter, 
especially within the 8 year MACHO dataset.   From internal MACHO comparisons, 
the error in period is $\rm \sigma_P = 0.000054^d$ \citep{minniti96}.

The observed $V$-amplitude of the RR Lyrae stars used in this analysis is found 
by performing an
eighth order Fourier decomposition to their light curves.  This in contrast
to the amplitudes listed by \citet{kunder08} which are derived from
a fourth order Fourier decomposition.  However, an eighth 
order Fourier fit better represents the minimum and maximum extremes of the 
RR Lyrae light curve, which is used to determine amplitude.  
The fourth order amplitudes begin
to deviate from eighth order amplitudes at $A_V$ $<$ 1.0 mag and are on average
$\sim$ 0.02 to 0.03 mag larger than the 8th order amplitudes.

A check on the accuracy of the MACHO RR0 Lyrae amplitudes is performed by 
using the LMC RR0 Lyrae stars in the OGLE-II and MACHO databases.  
Both surveys list RR0 Lyrae $V$-amplitudes.  

From a sample of 467 OGLE-II and
MACHO LMC RR0 Lyrae stars, the mean difference between
the OGLE $V$-amplitudes and the MACHO $V$-amplitudes is 0.05$\pm$0.13 mag,
where 0.13 refers to the dispersion about the mean.  The distribution is roughly Gaussian.  
As the dispersion is quite a bit
larger than the mean, the MACHO and OGLE $V$-amplitudes agree 
reasonably well with each other.  There is likely error in the OGLE $V$-amplitudes
as well as the MACHO $V$-amplitudes, so the random error in MACHO $V$-amplitude
is 0.10 mag, and
the zero-point in MACHO $V$-amplitude is up to 0.05 mag.  To diminish the
random error, much of the analysis is presented in the form of averages.

The MACHO bulge RR Lyrae light curves are of better quality than the MACHO LMC RR Lyrae light curves.  The LMC RR Lyrae light curves exhibit more scatter than the bulge RR Lyrae stars.
The LMC RR Lyrae stars are considerably fainter, with an average brightness of $V$=19.4 
and a mean error of a single point on the light curve is $\pm$0.07 mag
\citep{alcock96}.  This is in contrast to the bulge RR Lyrae stars which
have magnitudes that peak at $V$=15 mag
(Alcock et~al. 1997) and the mean error of a single point on the light
curve is $\pm$ 0.02 mag.  As a result, the discussion in the previous paragraph
likely overestimates the $V$-amplitude error.

\section{Period-Amplitude Plane}
\citet{sandage82} showed that a better parameter for the Oosterhoff 
classification is not the average of all the RR0 Lyrae periods but the 
measurement of, in the period-blue amplitude relationship, the difference 
between the average periods of an M3 RR Lyrae and the observed period at a 
fixed blue amplitude $A_B$.  
This is given by
\begin{equation}
\Delta \mbox{log} P = 0.129 A_B + 0.088 - \mbox{log} P(observed).
\end{equation}
Since then, this period shift has also been defined as a function of 
$V$-amplitude.  The least-squares fit to RR0 stars in the OoI prototype 
GC M3 is
\begin{equation}
\label{oo1pc}
\Delta \mbox{log} P = -0.14 A_V - 0.12 - \mbox{log} P(observed)
\end{equation}
and the least-squares fit to RR0 stars in the OoII GC M9 is
\begin{equation}
\Delta \mbox{log} P = -0.20 A_V + 0.026 - \mbox{log} P(observed)
\end{equation}
\citep{clement99}.

\begin{figure}[htb]
\includegraphics[width=16cm]{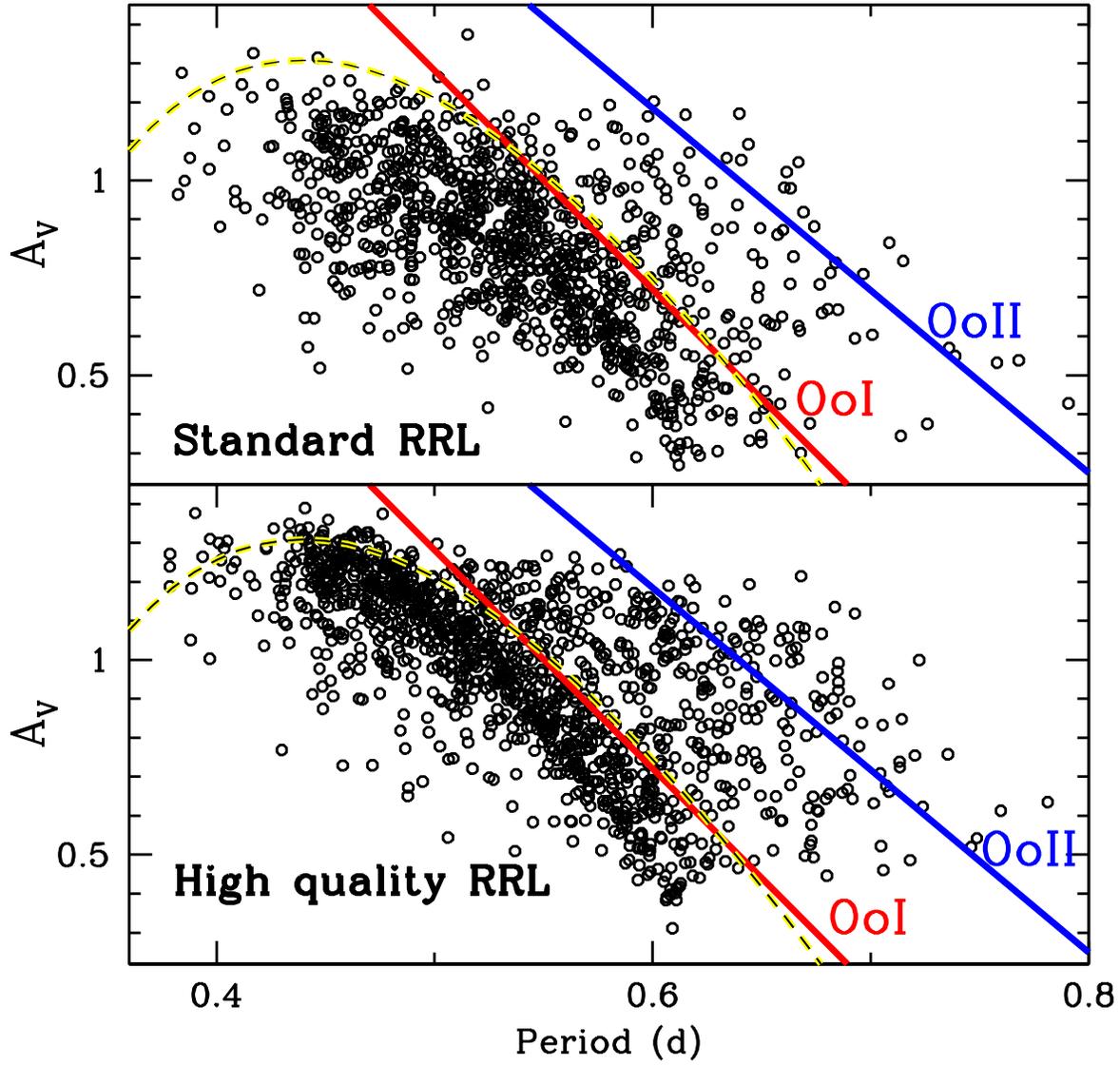}
\caption{\textit{Top:} Period-amplitude relation for the 1028 standard MACHO Bulge 
RR0 Lyrae variables with ``normal'' light curves.
\textit{Bottom:} Same as top panel but with the 1323 stars with high quality bulge light curves.
The straight lines were derived by \citet{clement99} from a least-squares 
fit to the principal
sequence of regular RR0 Lyrae stars in the OoI cluster M3 and 
OoII cluster M9.  The dashed line is a quadratic $\rm PA_V$
relation from \citet{cacciari05} derived from M3 RR0 Lyrae variables. 
\label{f2}}
\end{figure}

Figure~\ref{f2} shows the Period-Amplitude ($\rm PA_V$) diagram of the 1323
high quality MACHO bulge RR0 Lyrae variables in the bottom panel and the 1028
standard MACHO bulge RR Lyrae in the top panel.  Overplotted are the lines
that \citet{clement99} derived for Oosterhoff II and Oosterhoff I RR0 stars.  
In both panels it is clear that the majority of MACHO bulge field RR Lyrae do not fall along 
either of these lines, nor is there a "gap" between these lines.
The amplitudes would have to be shifted by $\sim$ 0.2 mag
for the main locus of stars to follow the OoI line.  From the above 
amplitude analysis, it is unlikely for the MACHO $V$-band amplitudes 
to be this discrepant.  There is considerable scatter in this plot, although the main
locus of the high quality RR Lyrae shows less scatter.  The
amplitude of light variation for Blazhko stars varies over timescales 
longer than the basic pulsation period and introduce scatter into the $\rm PA_V$
diagram if not identified.  Since only RR0 Lyrae stars with ``normal'' light curves are 
used here, there should not be many Blazhko variables introducing scatter to 
this plot.  Further,  only those with no clear indication of the Blazhko effect
were selected to be part of the high quality MACHO RR Lyrae.  This strengths
the result that their is no gap in the Bulge field RR0 Lyrae stars.

It is notable that the $\rm PA_V$ relation for the stars with shorter periods 
(\textit{i.e.} OoI characteristics), appear to follow a quadratic relation more 
than a linear relation.  This was also pointed out by \citet{cacciari05} with 
their study of the OoI characteristic M3 RR Lyrae stars.  Their quadratic $\rm PA_V$
relation is not used here because it does not give a solution for the 
period of an RR Lyrae star with an $A_V$ greater than 1.31 mag, and there are
RR0 Lyrae stars with a $V$-band amplitude greater than 1.31 mag in this sample.
As seen in Figure~\ref{f2}, the main locus of MACHO RR Lyrae stars does not fall 
on this line either.  There are $\sim$20 variables with $V$-band amplitudes 
greater than 1.31 mag in this sample.  This again illustrates the differences
between the properties of the RR Lyrae variables in GCs compared to those
located in the Galactic bulge.

\section{\dellp As An Oosterhoff Indicator}
The period shift defined by Equation~\ref{oo1pc}, \dellpc, is usually
made between the RR0 Lyrae stars in different globular clusters
\citep{sandage81,sandageea81, carney92, sandage93}.  
This quantity can be used as an
Oosterhoff indicator.  For example, when finding the period shift with 
respect to the OoI line, a period shift of \dellp $=$ 0 would 
indicate an Oosterhoff type I star, and a period shift of 
\dellp $\sim$ $-$0.07 would indicate an Oosterhoff type II star.  
Those values in-between show the
degree of deviation of an RR Lyrae star from an Oosterhoff class.

\begin{figure}[htb]
  \includegraphics[width=16cm]{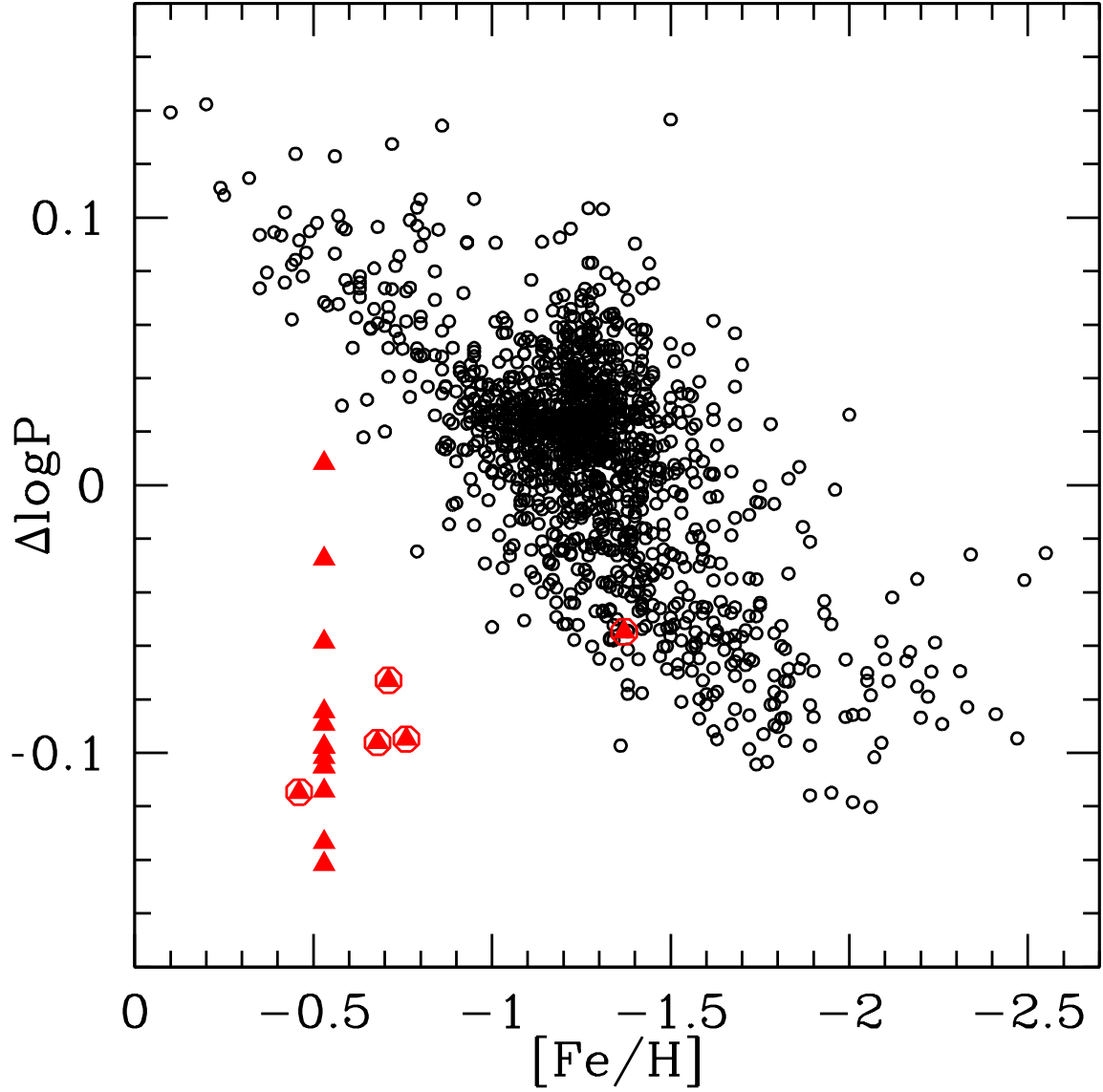}
  \caption{\dellp as a function of \feh for the Bulge field RR0 Lyrae stars with high
  quality light curves.  The red triangles represent RR Lyrae stars in the metal-rich 
  globular cluster NGC 6441with photometry from \citet{pritzl03}.  The red circled triangles 
  represent those stars with spectroscopic metallicities from \citet{clementini05}.
\label{f3}}
\end{figure}

Suntzeff, Kinman \& Kraft (1991) plot the period shift, \dellpc, 
with respect to the OoI line against metallicity using a sample of 171
RR0 field stars with spectroscopically determined metallicities.  These stars
are in the galactocentric distance range 4 - 30 kpc and are at located from
$(l,b)$ = (1 to 182$^\circ$, $-$18 to +89$^\circ$).
They found a gap at \dellp $\sim$ $-$0.03, and hence a clear separation 
between field stars of Oosterhoff type I and II.  Figure~\ref{f3} 
shows the period shifts of the high quality MACHO RR0 Lyrae stars as a function of 
their photometric metallicities.  There is no evidence of a gap. 

It is striking how many RR Lyrae stars
in the bulge sample extend to \dellp $>$ 0.05, a property
not common in RR0 Lyrae stars in globular clusters.  These stars 
are on average more metal-rich than the rest of the sample, and such 
metal-rich globular clusters are scarce
in the Milky Way.  The RR0 Lyrae stars in the metal-rich globular clusters, 
NGC 6441 and NGC 6388 (with \feh $\sim$ $-$0.6) have \dellp 
values that fall in the OoII region, with \dellp values of $\sim$ $-$0.09 days
(despite their high metallicity).
The \dellp values for these variables are plotted in Figure~\ref{f3}.
It is clear that the metal-rich RR Lyrae field stars in the bulge have quite different 
\dellp values than that of the metal-rich globular cluster RR Lyrae stars.  
NGC 6388 and NGC 6441 have been found to 
be more consistent with Oosterhoff type II clusters than Oosterhoff 
type I clusters.  However, for the metal-rich stars in the Galactic bulge, 
this is not the case.  This strongly suggests that the metal-rich RR Lyrae 
stars in the Galactic bulge may have a different origin than that in the 
metal-rich globular clusters. 

\begin{figure}[htb]
  \includegraphics[width=16cm]{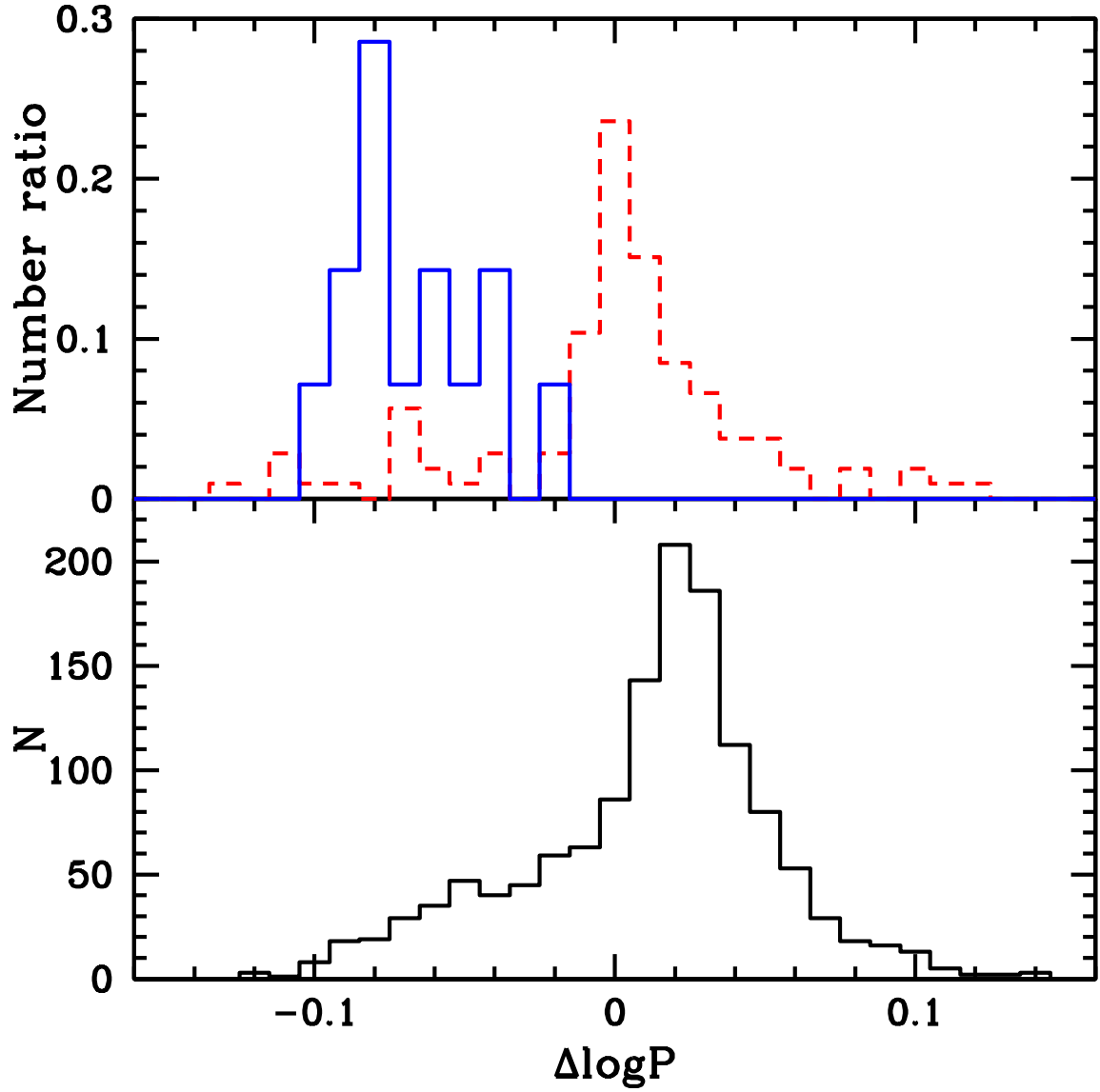}
  \caption{\textit{Bottom:} The histogram of \dellp from the high quality light curve RR Lyrae stars
  with respect to the OoI cluster M3.
  \textit{Top:} The histogram of RR0 Lyrae stars in the OoI globular clusters M3 and M5 (dashed),
  and the OoII clusters M2 and M9 (solid).
\label{f4}}
\end{figure}

In Figure~\ref{f4}, the distribution of the period shift, \dellpc, at fixed
$V$-band amplitude from the OoI line is shown.  Only the RR Lyrae
with high quality light curves are used here.  However, this histogram looks
almost identical if all the MACHO bulge stars are used.  
Also shown are the \dellp value from the OoI clusters M3 and M5, as well as 
the \dellp values from the OoII cluster M2 and M9.  The bulge RR0 Lyrae stars
have \dellp distributions that clearly do not coincide with
either OoI or OoII distributions.  This is not surprising, as this
was evident in the PA-plane as well.  The main locus of the 
distributions are shifted by \dellp $\sim$ 0.02 days relative to the OoI 
and OoII designations.   

Our survey of bulge RR Lyrae stars have found that the stars do not fit into the classic
OoI and OoII designations (with a period shift of 0.025 days), and indeed there is no evidence for an Oosterhoff dichotomy in our work.  In contrast, almost all of the recent surveys of field RR
Lyrae stars, i.e., LONEOS \citep{miceli08},  
SDSS-II Supernova Survey \citep{delee08}, ASAS \citep{szczygiel09}, have found that the
majority of their RR Lyrae \dellp values correspond well to the OoI
\dellp peak.  Moreover, these surveys have also identified a displaced secondary peak
corresponding to an OoII component.  The QUEST-I
sample \citep{vivas04} and the NSVS sample \citep{kinemuchi06}, however,
find no gap in their PA-diagram distributions.  \citet{szczygiel09} shows that
the NSVS $V$-amplitudes are incorrect, which had the
effect of masking the two Oosterhoff components in their sample.  THE QUEST 
RR Lyrae catalog, which did not find a Oosterhoff dichotomy in their
sample, probes deeper than the LONEOS survey (by $\sim$25 kpc),
and hence would give a more complete picture of the Galactic halo.  However, 
OoII RR Lyrae stars in general are located closer to the Galactic Plane
\citep{lee99, delee08}, and hence the depth of the QUEST survey may
have the effect of limiting the number of OoII RR Lyrae with respect to OoI
RR Lyrae.  None of these samples probe the Galactic 
bulge, and are located at quite 
different locations on the sky as the MACHO bulge RR0 Lyrae sample.

\subsection{\dellp and $V$-amplitude for a given period}
\citet{clement99} argue that for a given period, the $V$-amplitude
(a fairly good temperature indicator) for ``normal'' RR0 Lyrae stars does not 
depend on $\rm [Fe/H]$, but on Oosterhoff 
type.  They use seven globular clusters to come to this conclusion and 
apply the same compatibility test as \citet{kundchab08} to establish which 
cluster RR0 Lyrae stars had peculiar light curves.  \citet{clement99} then 
present evidence that the
Oosterhoff type of an RR Lyrae star is a function of its evolutionary state.  
If these arguments hold, \citet{clement99} discuss the implications on the 
determination of the ages of globular clusters from RR Lyrae stars.  We
use the Galactic bulge to investigate this notion.

\begin{figure}[htb]
  \includegraphics[width=16cm]{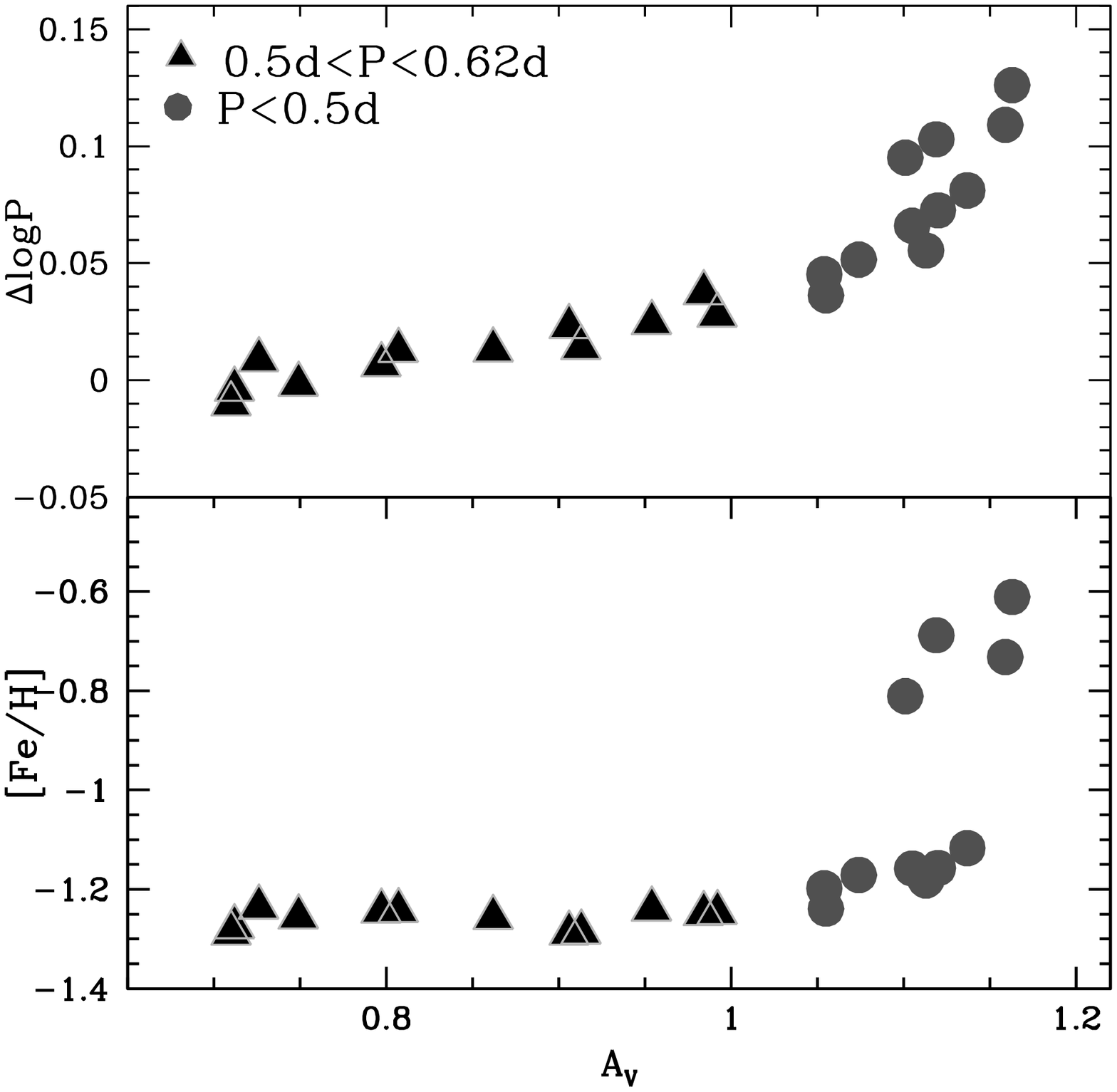}
  \caption{The Bulge RR0 Lyrae stars are binned in 0.01 day bins and the 
average $A_V$ in each period bin is plotted as a function of \feh (bottom)
and \dellp (top).
\label{f5}}
\end{figure}

The RR Lyrae variables are binned in 0.01 day bins and the average
\feh, $V$-amplitude and \dellp of the RR Lyrae in each period bin is found.
Figure~\ref{f5} shows the average $A_V$ in each period bin as a function of 
both \feh and \dellpc.  For clarity, only the RR0 Lyrae period bins with periods less 
than 0.62 day are shown.  It is clear from the bottom panel of Figure~\ref{f5}
that the average $V$-amplitude in each period
bin does not depend on metallicity.  In contrast, their is a slope in the $A_V$ 
and \dellp plane for the 
OoI-like RR0 Lyrae stars.  This slope is $\rm 0.11 \pm 0.02$ days/mag.  
With the exception for the most metal-rich stars, the 
stars with shorter periods (P$<$0.5d) show this result as well. 
Many studies have used the position of an RR Lyrae star 
in the period-$V$-amplitude diagram to estimate its metallicity 
\citep{alcock96, alcock00, kinemuchi06, sandage04, brown04}.
Here strong evidence is presented that cautions against using the period and 
$V$-amplitude of an RR Lyrae star to estimate its metallicity.  A similar 
analysis cannot be carried out with the OoII-like RR0 Lyrae stars, because 
they exhibit a wide range of metallicities with no clear function on 
$V$-amplitude for a given period.

\section{\dellp and Luminosity Differences}
\citet{lee99} performed an investigation on the luminosities of RR Lyrae 
variables 
with similar metallicities but different Oosterhoff classes.  They found
that RR Lyrae variables in the Oosterhoff II cluster M2 are intrinsically
brighter than those in Oosterhoff group I cluster M3, although these
clusters have similar \fehc.  Thus at least in this cluster pair, 
there is a discontinuity in the $\rm M^{\mathrm{RR}\,}_{V}$
-\feh relation.  Here a similar analysis is carried out using first
the bulge field RR0 Lyrae stars and second the local field
RR0 Lyrae stars.

\subsection{Bulge Field RR0 Lyrae Star Luminosities}
\begin{figure}[htb]
\includegraphics[width=16cm]{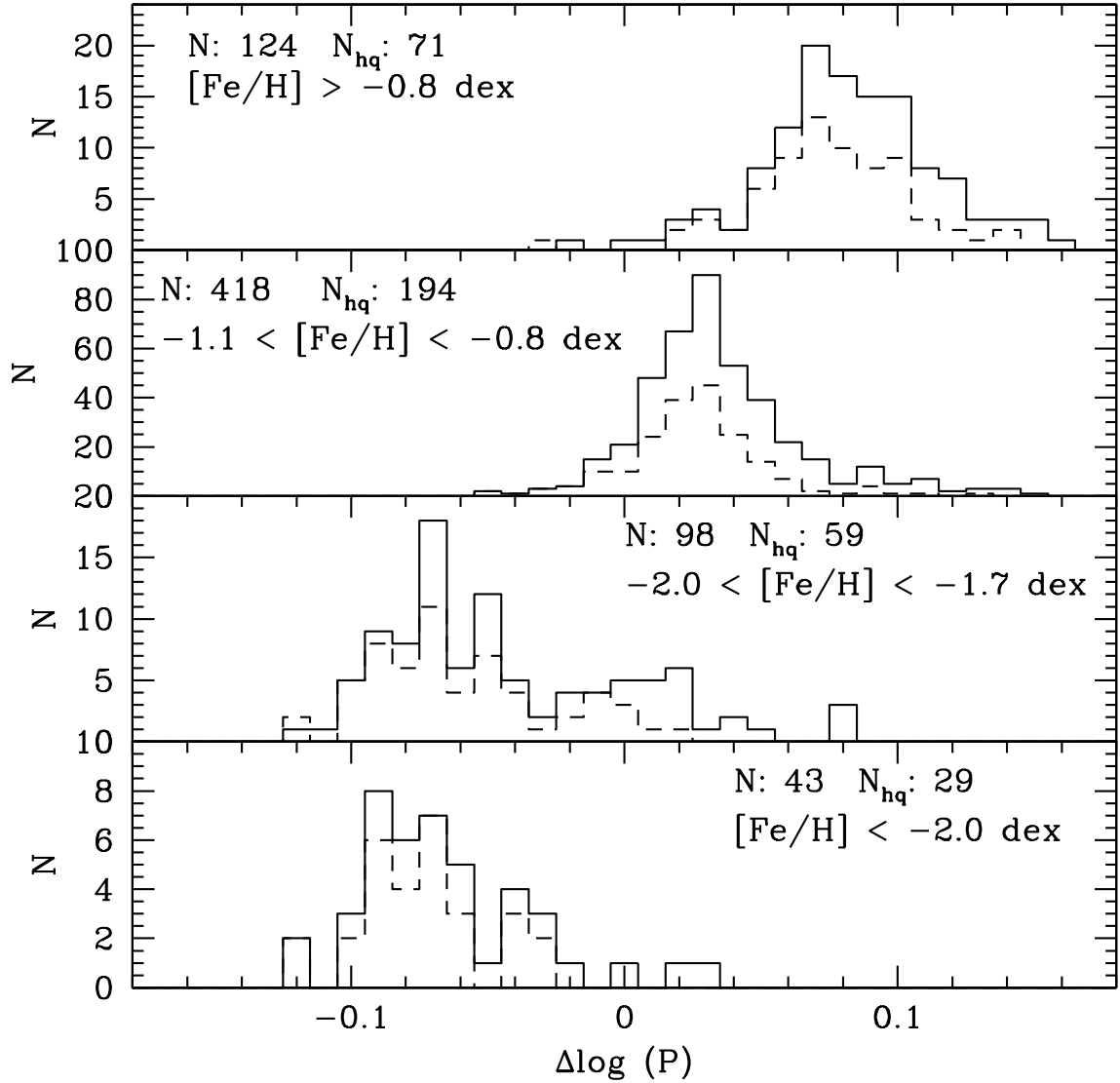}
\caption{Period shift distributions for the Bulge RR0 Lyrae field stars
with different metallicity ranges.  The dotted lines show the distribution
of the RR Lyrae with high quality light curves only.  The number of these
stars in each metallicity range is given by $\rm N_{hq}$, where as the total
number of stars in each metallicity range is given by N.
\label{f6}}
\end{figure}

Figure~\ref{f6} shows the \dellp distributions for the bulge
field RR0 Lyrae stars with different metallicity ranges.  The
high quality RR Lyrae stars are represented by the dotted lines,
and have very similar \dellp distributions to the complete sample.
Unlike for the most metal-rich RR0 Lyrae stars in the bulge field RR0 Lyrae
sample, the most metal-poor RR0 Lyrae samples have stars that occupy both 
the OoI and OoII regions as indicated by their \dellpc.  

\begin{figure}[htb]
\includegraphics[width=16cm]{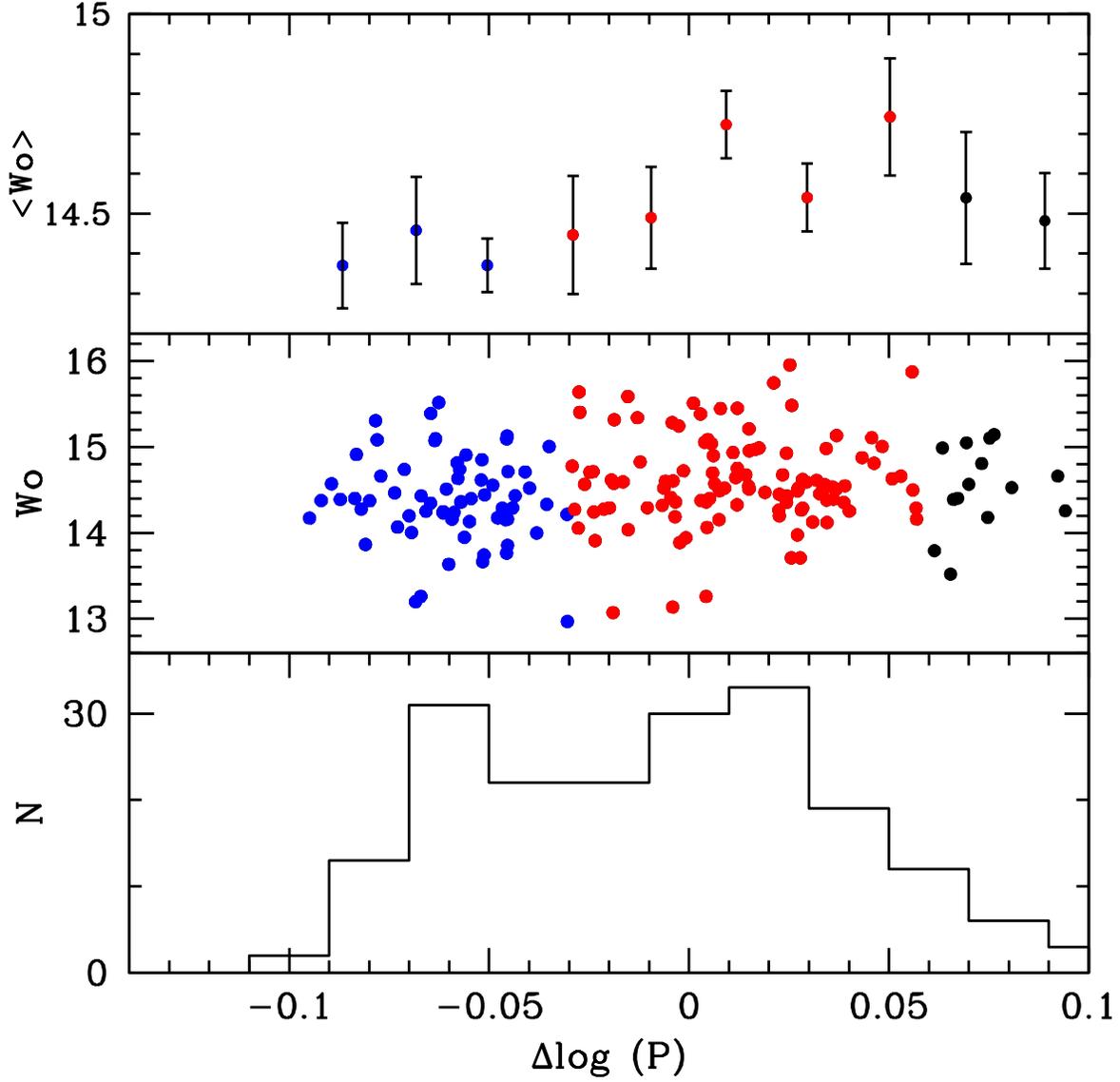}
\caption{Comparison of \dellp vs. the $\rm W_0$ for RR0 Lyrae stars
with similar \feh values.  \textit{bottom:} Histogram of the distribution
of the RR0 Lyrae stars with $-$1.7 $<$\feh$<$ $-$1.5 dex.   \textit{middle:}
Plot of the individual \dellp values as a function of $W_0$.
\textit{top:} Here the RR0 Lyrae stars are binned in 0.02 \dellp bins
and plotted against average $W_0$.
\label{f7}}
\end{figure}

The reddening-free apparent magnitude of the RR0 Lyrae variables with 
high quality light curves and photometric
metallicities of $-$1.7 $<$\feh$<$ $-$1.5 is shown in Figure~\ref{f7}
as a function of \dellpc.
The Wesenheit reddening free magnitude, $W_0$, is used to avoid
uncertain reddening corrections and is defined as $W_0 = V - 4.3(V-R)$, 
where the factor 4.3 is the selective extinction coefficient $R_{V,VR}$ derived
by \citet{kunder08}.  There are 107 stars with metallicities between 
\feh$=$$-$1.5 dex and \feh$=$$-$1.7 dex, and their distribution in the
\dellp plane is shown in the bottom panel of Figure~\ref{f6}.  It
is striking that this rather metal-poor sample has a \dellp distribution with 
an OoI-like tail.
Dividing the RR0 Lyrae at$-$0.03 \dellp days, the average $W_0$ for the
stars with \dellp $<$ $-$0.03 d, or the OoII-like stars, is 
$W_{0,II}$ = 14.33$\pm$0.07 mag.  The average $W_0$ for the
stars with 0.06 < \dellp $>$ $-$0.03, or the OoI-like stars, is 
$W_{0,I}$ = 14.58$\pm$0.07 mag.  No bias in location is found
for this particular sample of stars--$\rm W_0 $and \dellp is evenly 
distributed as a function of Galactic $l$ or $b$. This is consistent with 
\citet{kundchab08}, in which it is shown that for the entire sample 
of bulge stars, there is no strong evidence of $\rm W_0 $ as a function 
of Galacitic $l$ or $b$.  As there is no trend in \feh as a function
of \dellpc, this apparent magnitude difference shows that there is a discontinuity
in the absolute magnitude-\feh relation for this particular sample of stars.

\begin{figure}[htb]
\includegraphics[width=16cm]{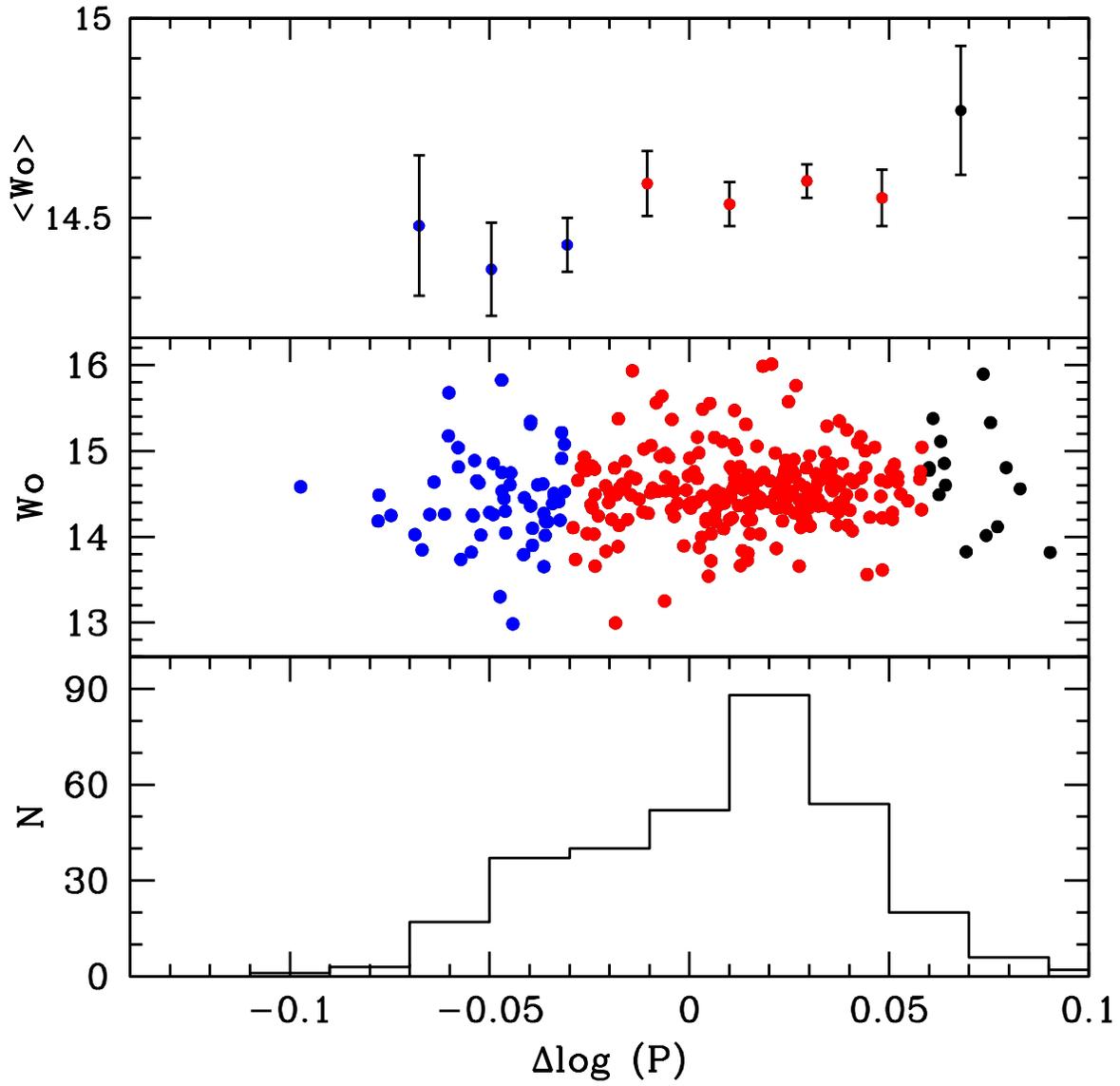}
\caption{Same as Figure~\ref{f6}, but for the 
metallicity distribution $-$1.5 $<$\feh$<$ $-$1.3 dex.
\label{f8}}
\end{figure}

This same analysis is carried out with the 320 bulge field RR0 Lyrae stars
that have metallicities in the $-$1.5 $<$\feh$<$ $-$1.3 dex range.
Again, only the stars with high quality light curves are used.
Figure~\ref{f8} shows how the reddening-free apparent magnitude 
decreases as the
RR0 Lyrae stars display more OoI-like \dellp values.  For this metallicity
distribution, there are a handful of stars at \dellp $>$ 0.06 days.
These stars were shown to have $V$-band amplitudes that were more similar 
to the $A_V$ found in the OoII-like stars rather than the $A_V$ found
in the OoI-like stars.  In Figure~\ref{f8} it is evident that their
$\rm <W_0>$ become more luminous than those OoI-like stars
at \dellp $<$ 0.06 days.   The average $W_0$ for those
stars with \dellp $<$ $-$0.03 days is $W_{0,II}$ = 14.44$\pm$0.07 mag.  
The average $W_0$ for the stars with \dellp $>$ $-$0.03 is 
$W_{0,I}$ = 14.56$\pm$0.03 mag.  Disregarding the stars with
\dellp $>$ 0.06 days, the average $W_0$ is $W_{0,I}$= 14.55$\pm$0.03 mag.  

The stars with high quality light curves in the $-$1.3 $<$\feh$<$ $-$1.1 dex
metallicity range show similar effects.  There are not enough stars in the more 
metal-rich or more metal-poor metallicity ranges to perform such an analysis.  
Table~\ref{tablum} summarizes these results.  When using all 
the RR Lyrae variables and not just those with high quality light curves, 
the same effect is seen.  Further, when all the light curves are used, one more 
metal-rich and one more metal-poor range can be investigated.  Both of
these metallicity ranges show that the OoII-type Bulge stars have brighter 
reddening-free magnitudes than OoI-type RR Lyrae.  This indicates that
unless the RR0 Lyrae absolute magnitude relation includes a 
period-amplitude term, the relation can lead to uncertainties of up to 
0.2 mag in absolute magnitude.  Furthermore, a quadratic 
$\rm M^{\mathrm{RR}\,}_{V}$-\feh relation would not correct for this.

\subsection{Local Field RR0 Lyrae Star Luminosities}
The previous section suggests that \feh alone cannot account for absolute 
magnitudes of RR Lyrae variables, and that an RR Lyrae variables position
on the $PA_V$ diagram may influence the absolute magnitude.
How robust this result is for local RR Lyrae stars is also of interest. 
Recently \citet{kovacs03} used the observed $V$-band,  $(V-K)$  colors 
and radial velocity curves of local RR Lyrae stars to compute their
luminosity, effective temperature, and radius variations.  
The results of this Baade-Wesselink (BW) analysis is presented in his 
Table~1.  $V$-amplitudes from the literature are obtained for these same stars, 
(Sandage 2004, Jones, Carney \& Latham 1988, Kunder et~al.(2008),
Lub 1977) and their \dellp was computed.  \feh values for these
stars come from \citet{layden96}.

\citet{bono03} compiled a list of 36 RR0 Lyrae stars (field and GGC) for 
which accurate $V$ and $K$ light curves, reddening and metallicity estimates 
were known.  From these parameters and their updated pulsation models, 
the $M_K$ absolute magnitude was derived and in turn, the $M_V$ absolute
magnitude.  \dellp values for 15 stars listed in the \citet{bono03} 
table that are not present in the \citet{kovacs03} sample are obtained.

\begin{figure}[htb]
\includegraphics[width=16cm]{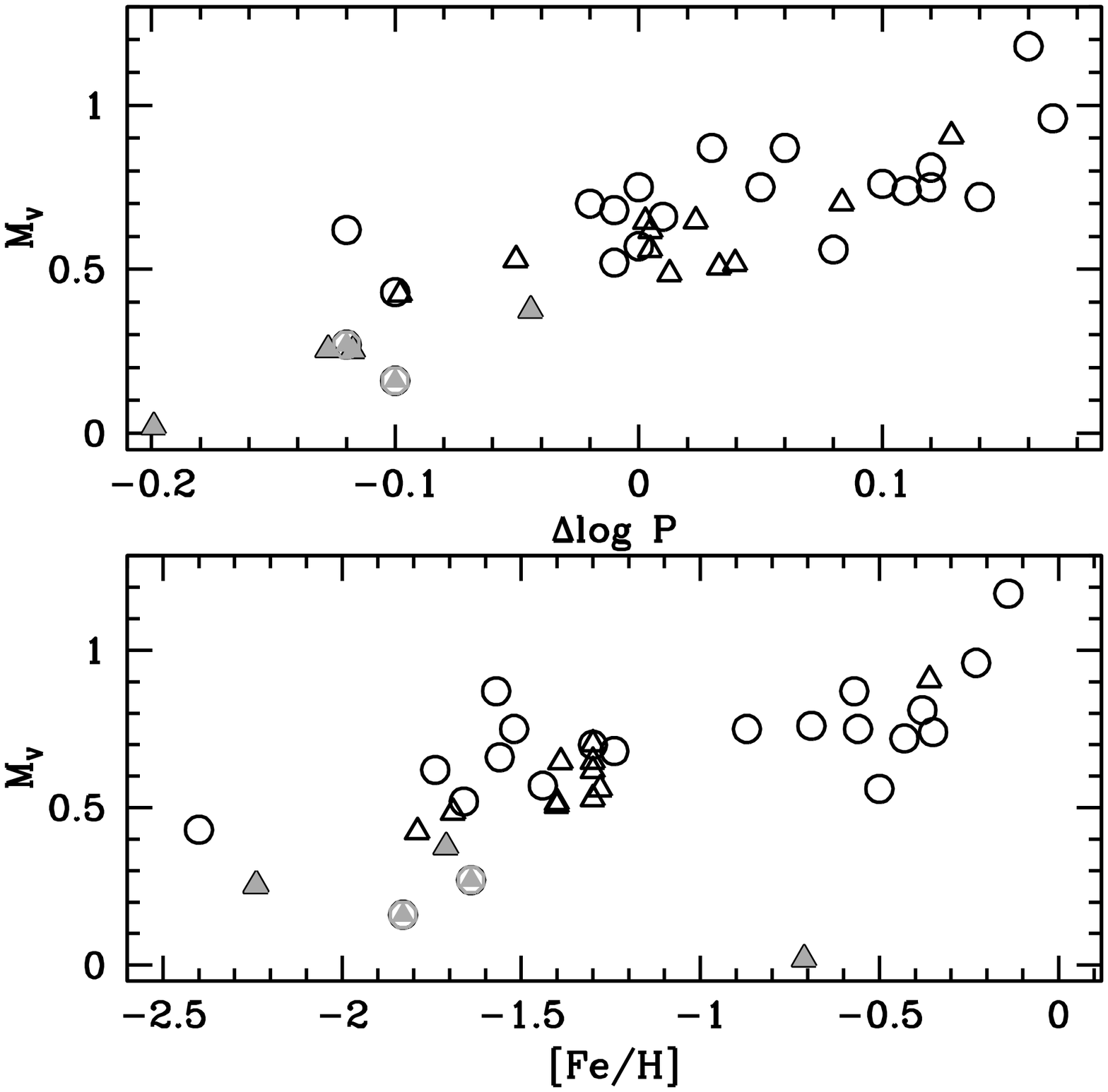}
\caption{A plot of local field RR0 Lyrae stars with $\rm M^{\mathrm{RR}\,}_{V}$
values derived from a Baade-Wesselink analysis,
as a function of \feh and \dellpc.  The open circles indicate those
stars with a Baade-Wesselink $\rm M^{\mathrm{RR}\,}_{V}$
from \citet{kovacs03} and the open triangles indicate those
stars with an $\rm M^{\mathrm{RR}\,}_{V}$ from \citet{bono03}.  The filled
triangles highlight those stars with the brightest absolute magnitudes, and
seem to be the most discrepant in the $\rm M^{\mathrm{RR}\,}_{V}$ - \feh plot.
\label{f9}}
\end{figure}

Figure~\ref{f9} shows the absolute magnitude
of 36 resulting RR Lyrae stars as a function of \feh and 
\dellpc.  \citet{kovacs03} and \citet{bono03} do not list an error in 
$\rm M^{\mathrm{RR}\,}_{V}$, but the standard methodology of 
the BW analysis typically results in errors in absolute magnitude
of 0.2 mag \citep{liu90}.  It is also evident with the local field RR Lyrae
stars with \dellp $< -$0.03 days, (\textit{i.e.} the OoII-like
stars) have on average brighter absolute magnitudes.  

There is a cluster of seven stars with \feh of $\sim$ $-$1.75 dex, and
these stars have a range in absolute magnitudes of 0.6 mag.  This 
is indication that a second parameter may influence the absolute
magnitude of an RR Lyrae.
It is especially striking that the five RR Lyrae stars with
the brightest absolute magnitudes, represented in Figure~\ref{f9}
by filled triangles, introduce large scatter in the \feh vs 
$\rm M^{\mathrm{RR}\,}_{V}$ plot, but follow the 
\dellp versus $\rm M^{\mathrm{RR}\,}_{V}$ trend reasonably well.
These stars especially highlight that an RR Lyrae variable's 
metallicity may not always indicate its absolute magnitude.
When determining absolute magnitude, the stars'
\dellp must be taken into account.  

A least-squares fit to \feh versus $\rm M^{\mathrm{RR}\,}_{V}$ results in
a dispersion about the fit of 0.19, and the least-squares fit of \dellp versus
$\rm M^{\mathrm{RR}\,}_{V}$ results in a dispersion about the fit of 0.13. 
A simultaneous fit of \feh and \dellp versus absolute magnitude also
results in a dispersion about the fit of 0.15.  The BW stars 
have \dellp values that are tightly correlated with \feh and hence it is 
difficult to disentangle the effect of absolute magnitude versus \dellp 
and \fehc.

Although most RR Lyrae variables are found in relatively
metal-poor globular clusters, two metal-rich GCs, 
NGC 6388 and NGC 6441, have been found to host
RR Lyrae variables.  These clusters have very 
unusual HB morphologies for their metallicities \citep{rich97}.
Although NGC 6388 and NGC 6441 have metallicities of 
\feh $\sim$ $-$0.6 dex, 
which are indicative of OoI GCs, their \dellp values fall
inline with the OoII GCs, with \dellp $\sim$ $-$0.09 days.
Traditionally, their \feh would indicate that these stars are
intrinsically fainter than most other RR Lyrae variables.
However, models introduced by \citet{sweigart98} require that the 
HBs of these clusters be unusually bright for the metal 
abundances of the clusters.  
Figure~\ref{f8} shows that these stars should be
intrinsically \textit{brighter} than most RR Lyrae variables,
because of their values of \dellpc.

\section{Conclusion}
Field bulge RR0 Lyrae stars are used to search for evidence of the
Oosterhoff dichotomy within the field population in the Milky Way halo.
The period-$V$-amplitude plane for the bulge field RR0 Lyrae stars does
not have a locus that matches either the OoI or OoII period-$V$-amplitude
plane of the MW globular clusters.  The bulge field RR0 Lyrae stars
are in general more metal-rich than even the OoI-type GCs. 
Consequently, their main locus is shifted in the same manner 
as the OoI-type GCs
is shifted compared to the more metal-poor OoII-type GCs.

The metal-rich bulge RR Lyrae variables lie on a different
period-$V$-amplitude plane than the metal-rich GC RR0 Lyrae stars
in NGC 6388 and 6411.  The metal-rich GC RR0 Lyrae stars have $PA_V$
properties more similar to the OoII-type GC RR0
Lyrae stars, whereas this is reversed for the metal-rich bulge
RR Lyrae variables.  It is unlikely that the metal-rich RR Lyrae stars
in the Galactic bulge and the metal-rich RR Lyrae stars in NGC 6388
and NGC 6411 come from a similar origin.

Although the period-amplitude relations for OoI and OoII-type
GCs use linear forms to determine the period shift, \dellpc, of
a GC, the period-$V$-amplitude plane of the thousands of
Galactic bulge RR0 Lyrae stars show that a quadratic
 period-$V$-amplitude relation would be a better approximation.
 \citet{cacciari05} found this with RR0 Lyrae variables in M3.  However,
their relation does not give a solution for the period of an RR Lyrae star with 
a $V$-amplitude greater than 1.31 mag.  There are bulge field
RR0 Lyrae stars with a $V$-band amplitude greater than 1.31 mag.

The reddening-free apparent magnitude differences between the OoI and 
OoII-type bulge field RR Lyrae stars are discussed.  As was found in RR0 Lyrae
stars in GCs \citep{lee99}, RR0 Lyrae variables with
OoII-type properties are about 0.2 mag brighter than those with
OoI-type properties.  The differences in the evolutionary stages 
of the OoI and OoII-type variables are probably the cause for
this luminosity difference.  A quadratic dependence of  \feh - 
$\rm M^{\mathrm{RR}\,}_{V}$ would not correct this effect.  
Using local field RR Lyrae stars with absolute magnitude values
determined from a Baade-Wesselink analysis, it is shown
that \dellp is a better indicator of an RR0 Lyrae stars' 
absolute magnitude than \fehc.

\clearpage

\begin{scriptsize}
\begin{sidewaystable}[p]\small
\caption{Apparent Magnitude Differences Between Oosterhoff I-like and Oosterhoff II-like bulge RR0 Lyrae Field Stars }
\label{tablum} \centering
\begin{tabular}{l c c c c c c c c } \\ \hline
Metallicity Bin & $\rm <W_{0,II}>$ & $\rm <W_{0,I}>$ & $\rm <\feh_{II}>$ & $\rm <\feh_I>$ & $\rm <P_{II}>$ & $\rm <P_{I}>$ & $\rm <A_{V_{II}}>$ & $\rm <A_{V_{I}}>$  \\ \hline
$-$1.7 $<$\feh$<$ $-$1.5 & 14.33$\pm$0.07 & 14.58$\pm$0.07 & -1.60$\pm$0.01 & -1.58$\pm$0.01 &  0.62$\pm$0.01 & 0.53$\pm$0.01 & 1.02$\pm$0.02 & 1.04$\pm$0.03 \\
$-$1.5 $<$\feh$<$ $-$1.3 & 14.44$\pm$0.07 & 14.55$\pm$0.03 & -1.40$\pm$0.01 & -1.37$\pm$0.01 & 0.62$\pm$0.01 & 0.52$\pm$0.00 & 0.95$\pm$0.03 & 1.04$\pm$0.01 \\
$-$1.3 $<$\feh$<$ $-$1.1 & 14.33$\pm$0.06 & 14.53$\pm$0.02 & -1.21$\pm$0.01 & -1.21$\pm$0.00 & 0.63$\pm$0.01 & 0.53$\pm$0.00 & 0.84$\pm$0.03 & 0.98$\pm$0.01 \\
\hline
\end{tabular}
\end{sidewaystable}
\end{scriptsize}

\clearpage

\clearpage


\clearpage

\begin{thebibliography}{}
\bibitem[Alcock et~al.(1996)]{alcock96}Alcock, C., Allsman, R.A., Axelrod, T.S., Bennett, D.P.\
Cook, K.H., Freeman, K.C., Griest, K., Marshall, S.L, Peterson, B.A., Pratt, M.R., Quinn, P.J.,
Rodgers, A.W., Stubbs, C.W., Sutherland, S., \& Welch, D.L.\ 1996, AJ, 111, 3
\bibitem[Alcock et~al.(1997)]{alcock97} Alcock, C. et al. 1997, ApJ, 486, 697
\bibitem[Alcock et~al.(2000)]{alcock00} Alcock, C., et al.\ 2000, AJ, 119, 2194
\bibitem[Brown et~al.(2004)]{brown04} Brown, T. M., Ferguson, H., Smith, E., Kimble, R. A., Sweigart, A. V., Renzini, A., \& Rich, R. M. 2004, AJ, 127, 2738
\bibitem[Bono et~al.(2003)]{bono03} Bono, G., Caputo, F., Castellani, V., Marconi, M., Storm, J., 
\& Degl'Innocenti, S.\ 2003, MNRAS, 344, 1097
\bibitem[Bono et~al.(1995)]{bono95}Bono, G., Caputo, F. \& Marconi, M.\ 1995, AJ, 110, 2365
\bibitem[Bono et~al.(2007)]{bono07} Bono, G., Caputo, F., di 
Criscienzo, M.\ 2007, A\&A, 476, 779
\bibitem[Cacciari et~al.(2005)]{cacciari05}Cacciari, C., Corwin, T. M., 
\& Carney, B. W.\ 2005, AJ, 129, 267
\bibitem[Carney et~al.(1992)]{carney92}Carney, B. W., Storm, J., \& Jones, 
R. V. 1992, ApJ, 386, 663
\bibitem[Carretta et~al.(2000)]{carretta00} Carretta, R.G., Gratton, G., Clementini, F., Fusi Pecci, ApJ, 533, 215 
\bibitem[Catelan(2005)]{catelan05} Catelan, M. 2005, in Resolved Stellar 
Populations, ASP Conf. Ser., ed. D. Valls-Gabaud \& M. Ch\'avez 
(San Francisco: ASP), in press
\bibitem[Catelan(2007)]{catelan07} Catelan, M.\ 2007, in Globular Clusters: Guides to Galaxies, ed. D. Geisler \& T. Richtler (ESO; Berlin: Springer)
\bibitem[Clement \& Shelton(1999)]{clement99} Clement, C.M. \& Shelton, I. 1999, \apj, 515, 88
\bibitem[Clementini et~al.(2005)]{clementini05} Clementini, G. Gratton, R.
Bragaglia, A., Ripepi, V., Fiorenzano, A.F.M., Held, E.V. \& Carretta, E.\
2005, ApJ, 630L, 145
\bibitem[de Lee(2008)]{delee08} de Lee, N. 2008, Ph.D. thesis, 10, Michigan State University
\bibitem[Eggen et~al.(1962)]{eggen62}Eggen, O. J., Lynden-Bell, D., \& Sandage, A. R. 1962, ApJ, 136, 748
\bibitem[Jones et~al.(1988)]{jones88} Jones, R.V., Carney, B.W. \& Latham, D.W.\ 1998, ApJ, 332, 206
\bibitem[Jurcsik \& Kov\'{a}cs(1996)]{jk96} Jurcsik, J., \& Kov\'{a}cs, G.\ 1996, \aap, 312, 111
\bibitem[Kinemuchi et~al.(2006)]{kinemuchi06} Kinemuchi, K., Smith, H.A.,
Wo\'{z}niak, P.R., McKay, T.A., \& The ROTSE Collaboration\ 2006, AJ, 132, 1202
\bibitem[Kov\'{a}cs \& Kanbur(1998)]{kovkan98} Kov\'{a}cs, G., \& 
Kanbur, S. M.\ 1998, MNRAS, 295, 834
\bibitem[Kov\'{a}cs (2003)]{kovacs03} Kov\'{a}cs, G.\  2003, MNRAS, 342, 58
\bibitem[Kunder et~al.(2008)]{kunder08}Kunder, A.M., Popowski, P., Cook, K.H., \& Chaboyer, B.\  2008, AJ, 135, 631
\bibitem[Kunder \& Chaboyer(2008)]{kundchab08}Kunder, A.M. \& Chaboyer, B.\  2008, AJ, 135, 631
\bibitem[Layden et~al.(1996)]{layden96} Layden, A.C., Hanson, R.B., Hawley, S.L., Klemola, A.R. \& Hanley, C.J.\ 1996, AJ, 112, 5 
\bibitem[Lee \& Carney(1999)]{lee99}Lee, J. \& Carney, B.W.\ 1999, AJ, 118, 1373
\bibitem[Lee et~al.(1990)]{lee90}  Lee, Y.-W., Demarque, P., \& Zinn, R.\ 1990, ApJ, 350, 155
\bibitem[Liu \& Janes(1990)]{liu90} Liu, T. \& Janes, K.A.\ 1990, ApJ, 360, 561
\bibitem[Lub(1977)]{lub77} Lub, J.\ 1977, A\&AS, 29, 345
\bibitem[Minniti(1996)]{minniti96} Minniti, D.\  1996, ApJ. 459. 175
\bibitem[Miceli et~al.(2008)]{miceli08}Miceli, A., Rest, A., Stubbs, C.W., 
Hawley, S.L., Cook, K.H., Magnier, E.A., Krisciunas, K., Bowell, E., \& 
Koehn, B.\ 2008, ApJ, 678, 865
\bibitem[Oosterhoff(1939)]{ooster39} Oosterhoff, P. Th.\ 1939, Observatory
62, 104
\bibitem[Pritzl et~al.(2001)]{pritzl01} Pritzl, B. J., Smith, H. A., Catelan, M., \& Sweigart, A. V.\ 2001, AJ, 122, 2600
\bibitem[Pritzl et~al.(2003)]{pritzl03} Pritzl, B. J., Smith, H. A., 
Stetson, P.B., Catelan, M., Sweigart, A. V., Layden, A.C. \& Rich, R.M.\ 
2003, AJ, 126, 1381
\bibitem[Pritzl(2004)]{pritzl04} Pritzl, B. J., Armandroff, T. E., Jacoby, G. H., 
\& Da Costa, G. S.\ 2004, AJ, 127, 318
\bibitem[Rich et~al.(1997)]{rich97}Rich, R. M., et al. 1997, ApJ, 484, L25 
\bibitem[Sandage(1958)]{sandage58}Sandage A. 1958, ApJ, 127, 515 
\bibitem[Sandage(1981)]{sandage81}Sandage, A.\ 1981, ApJ, 248, 161
\bibitem[Sandage(1982)]{sandage82} Sandage, A.\ 1982, \apj, 252, 574
\bibitem[Sandage(1993)]{sandage93}Sandage, A.\ 1993, AJ, 106, 687 
\bibitem[Sandage(2004)]{sandage04} Sandage, A.\ 2004, AJ, 128, 858
\bibitem[Sandage(2006)]{sandage06} Sandage, A.\ 2006, AJ, 131, 1750
\bibitem[Sandage et~al.(1981)]{sandageea81}Sandage, A., Katem, B., \& 
Sandage, M. 1981, ApJS, 46, 41 
\bibitem[Sandage \& Tammann(2006)]{sandtamm06}Sandage, A., \& Tammann, G. A. 2006, 
ARA\&A, 44, 93
\bibitem[Searle \& Zinn(1978)]{searle78}Searle, L., \& Zinn, R. 1978, ApJ, 225, 357 
\bibitem[Siegel et~al.(2000)]{siegel00} Siegel, M. H., \& Majewski, S. R.\ 
2000, AJ, 120, 284 
\bibitem[Suntzeff et~al.(1991)]{suntzeff91}Suntzeff, N., Kinman, T, \& Kraft, R. 1991, 
ApJ, 367, 528 
\bibitem[Sweigart \& Catelan(1998)]{sweigart98} Sweigart, A. V., \& Catelan, M.\ 
1998, ApJ, 501, L63
\bibitem[Szczygiel et~al.(2009)]{szczygiel09} Szczygiel, D.M. (in prep.)
\bibitem[van Albada \& Baker(1971)]{vanalbada73} van Albada, T.S., \& Baker, N.,
1973, ApJ, 185, 477
\bibitem[Vivas et~al.(2004)]{vivas04}Vivas, A. K., et al. 2004, AJ, 127, 1158
\bibitem[Walker(2000)]{walker00}Walker, A.R.: in IAU Colloq. 176, The Impact of Large-Scale Surveys on Pulsating Star Research, ed. L. Szabados \& D. Kurtz. (San Francisco: ASP), 165 (2000) 


\end{thebibliography}
\end{document}